\documentclass{iaus}
\usepackage{graphicx}

\title[Nuclear Star Clusters in Spirals] 
{Nuclear Star Clusters (Nuclei) in Spirals and Connection to 
Supermassive Black Holes}

\author[van der Marel et al.]
{Roeland P.~van der Marel$^1$, Joern Rossa$^2$, Carl Jakob Walcher$^3$,
Torsten Boeker$^4$, Luis C.~Ho$^5$, Hans-Walter Rix$^6$, Joseph C.~Shields$^7$}

\affiliation{$^1$STScI, 3700 San Martin Drive, Baltimore, MD 21218, 
USA\\[\affilskip]
$^2$Dept.~of Astronomy, University of Florida, Gainesville, FL 32611, 
USA\\[\affilskip]
$^3$Laboratoire d'Astrophysique de Marseille, F-13376 Marseille Cedex 12, 
France\\[\affilskip]
$^4$European Space Agency, Department RSSD, 2200 AG Noordwijk, 
The Netherlands\\[\affilskip]
$^5$Carnegie Institution of Washington, 813 Santa Barbara Street, 
Pasadena, CA 91101, USA\\[\affilskip]
$^6$Max-Planck-Institut f\"ur Astronomie, K\"onigstuhl 17, 
D-69117 Heidelberg, Germany\\[\affilskip]
$^7$Dept.~of Physics and Astronomy, Ohio University, Athens, OH 45701, 
USA\\[\affilskip]}

\pubyear{2007}
\volume{241}
\pagerange{}
\date{}
\jname{Stellar Populations as Building Blocks of Galaxies} 
\editors{A. Vazdekis and R. Peletier., eds.}

\begin{document}

\maketitle

\begin{abstract}
HST observations have revealed that compact sources exist at the
centers of many, maybe even most, galaxies across the Hubble
sequence. These sources are called ``nuclei'' or also ``nuclear star
clusters'' (NCs), given that their structural properties and position
in the fundamental plane are similar to those of globular
clusters. Interest in NCs increased recently due to the independent
and contemporaneous finding of three groups (Rossa et al. for spiral
galaxies; Wehner \& Harris for dE galaxies; and C\^ot\'e et al. for
elliptical galaxies) that NC masses obey similar scaling relationships
with host galaxy properties as do supermassive black holes. Here we
summarize the results of our group on NCs in spiral galaxies. We
discuss the implications for our understanding of the formation and
evolution of NCs and their possible connection to supermassive black
holes.
\end{abstract}

\firstsection

\section{Introduction}
\noindent High spatial resolution observations with the Hubble Space
Telescope (HST) have revealed that many galaxies have a compact source
in their very center. These ``nuclei'' have structural properties that
are similar to those of globular clusters (see below), and they are
therefore also called ``nuclear star clusters'' (NCs). Figure~1a shows
the nearby Sd galaxy NGC~300 as an example. This galaxy has a NC,
which is easily identified as a separate component because of the
marked upturn in the surface brightness profile at small radii
(Figure~1b). Boeker et al.~(2002, 2004) observed a sample of 77
late-type spiral galaxies with HST/WFPC2 and found NCs in 77\% of the
sample. Fits to the two-dimensional images yield the effective radii
and luminosities of the NCs, which for late-type spirals have median
values $r_{\rm eff}= 3.5$ pc and $L_I = 10^{6.2} L_{\odot}$.

The study of NCs in late-type spirals is facilitated by the faintness
of an underlying bulge/spheroid component. The same is true for the
study of nuclei in dE galaxies. However, NCs are also found in other
galaxy types. Carollo et al.~(1997, 1998) report a detection frequency
of 50--60\% in early-type spirals using HST/WFPC2 imaging. C\^ot\'e et
al.~(2006) report the detection of nuclei in 66--82\% of
intermediate-luminosity Virgo cluster elliptical galaxies (but no
detections in the most luminous giant ellipticals) using HST/ACS
imaging. The detection of NCs does become progressively more difficult
as the density of the underlying spheroid increases, which is a
problem especially for elliptical galaxies with steep cusps. The
inferred NC characteristics then depend sensitively on the assumed
spheroid brightness profile (Lauer et al.~2006; Ferrarese et
al.~2006b).

\begin{figure}[t]
\includegraphics[width=0.25\hsize]{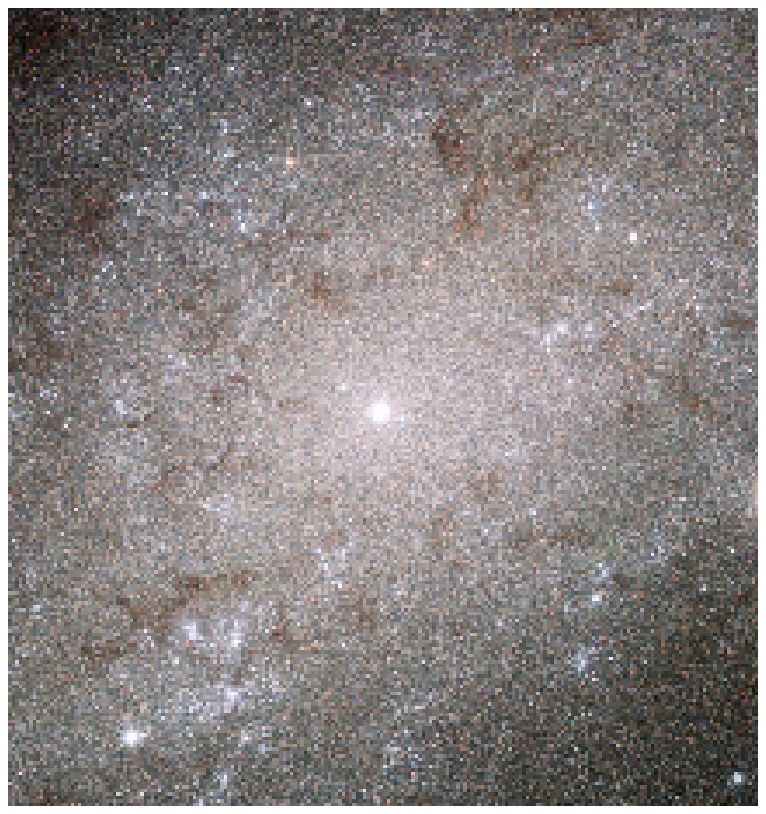}\qquad\hfill
\includegraphics[width=0.267\hsize]{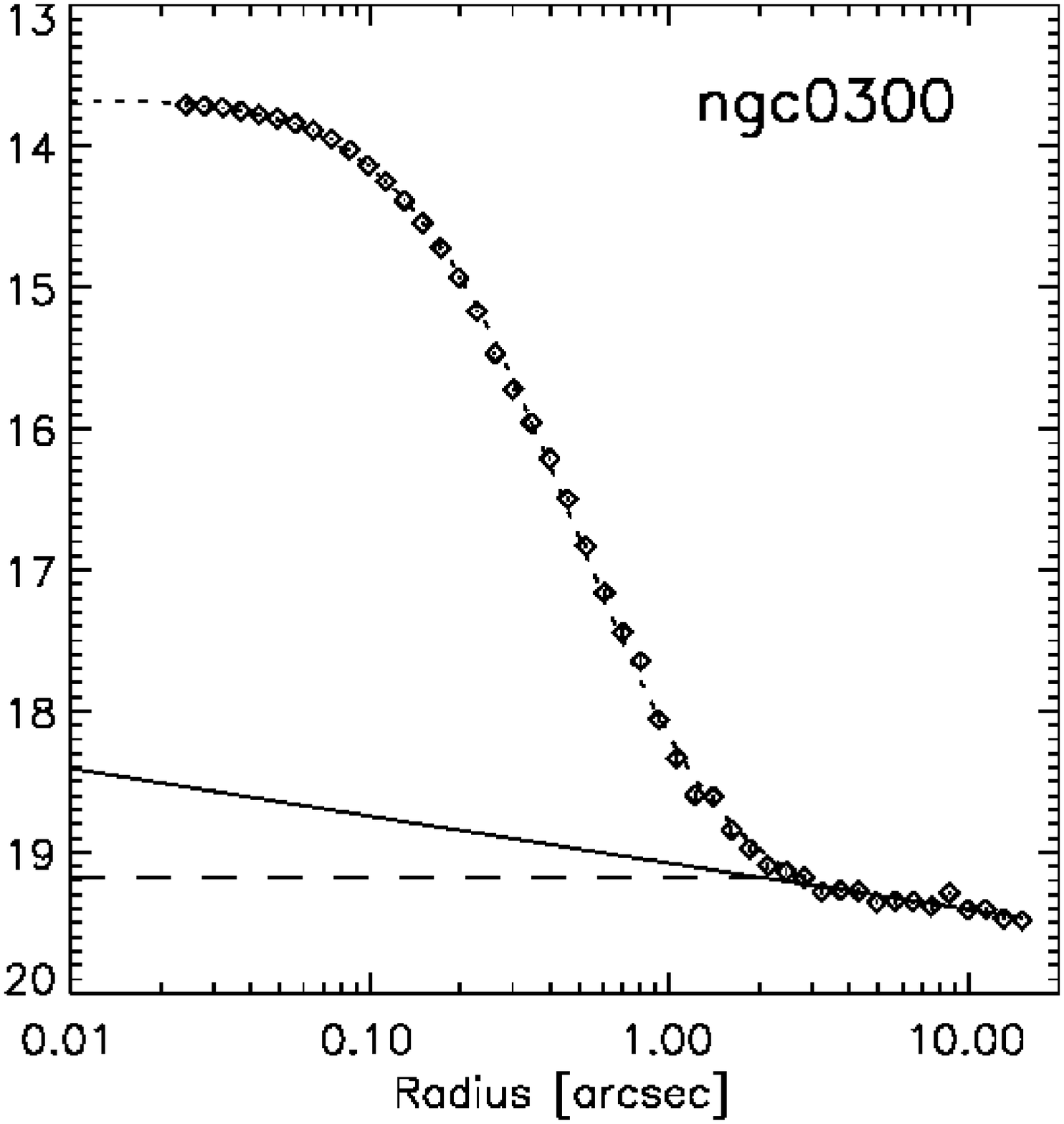}\hfill
\includegraphics[width=0.35\hsize]{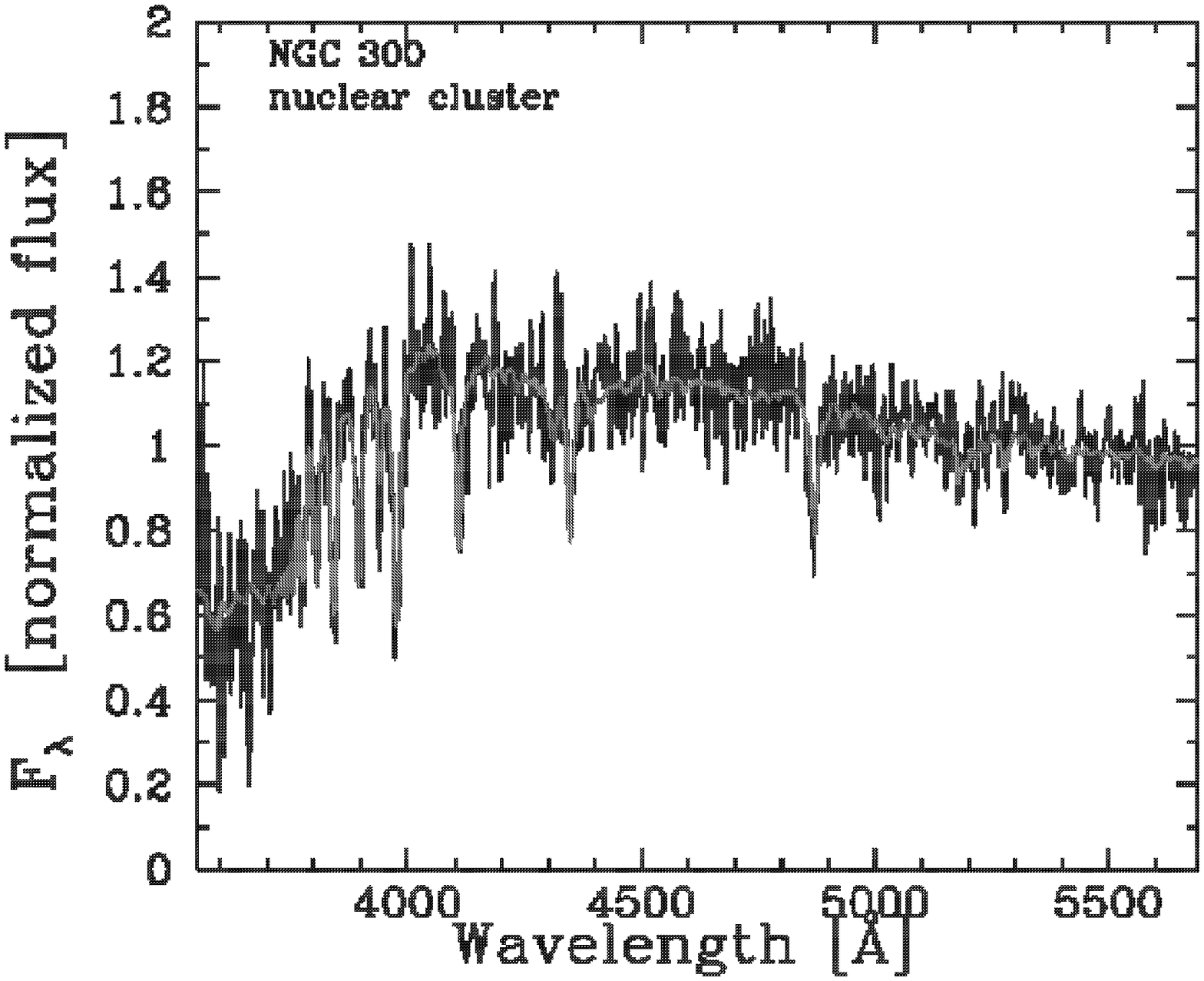}
\caption{(\textit{a}) HST/ACS color composite image of the central
$1.2 \times 1.2$ kpc of the Sd galaxy NGC 300 at $D \approx 1.8$ Mpc
(from the observations of Bresolin et al.~2005). A prominent nuclear
star cluster (NC) resides at the center of the galaxy.  (\textit{b})
$I$-band surface brightness profile in mag/arcsec$^2$ for the central
$\sim 10'' = 87$ pc, as determined from HST/WFPC2 data of the same
galaxy (Boeker et al.~2002). The NC is evident as a marked upturn in
the brightness profile, as compared to inward extrapolations (solid,
dashed lines) of the the brightness profile at large radii. King-model
fits to the two-dimensional image imply a NC luminosity $L_I =
10^{6.2} L_{\odot}$ and effective radius $r_{\rm eff} = 2.9$ pc
(Boeker et al.~2004). (\textit{c}) HST/STIS spectrum of the NC in
NGC~300 (from Rossa et al.~2006). The red curve shows the best
spectral population fit, which has $Z=0.004$, $A_V = 0.4$, a
luminosity-weighted mean $\langle \log ({\rm age/Gyr}) \rangle = 8.63$
and $M/L_B = 0.51 M_{\odot}/L_{B,\odot}$. The implied NC mass is $M =
10^{5.7} M_{\odot}$.}
\end{figure}

\section{Spectroscopy of Nuclear Star Clusters in Spiral Galaxies}
\noindent NCs/nuclei are a common component of galaxies across the
Hubble sequence. It is therefore important to have observational
knowledge of their stellar populations, ages, and masses. We pursued a
spectroscopic program to measure these quantities for spiral galaxies.
We used the UVES echelle spectrograph on the VLT to obtain spectra of
the NCs in nine late-type spirals (Walcher et al.~2005, 2006). The
high spectral resolution of these data allows the measurement of the
NC velocity dispersions from the near-IR Ca II triplet line
broadening. The results fall in the range $\sigma = 13$--34 km/s. When
interpreted with spherically symmetric dynamical models tailored to
fit the HST surface brightness profiles, these measurements yield the
NC masses. Figure~2a shows the NC properties in a face-on fundamental
plane projection. The NCs fall on a high-mass extension of the
globular cluster sequence.\looseness=-2

The VLT ground-based spatial resolution contaminates the NC light with
that of the underlying galaxy. This is manageable in the latest-type
spirals, but makes it difficult to study NCs in earlier-type
spirals. We therefore also pursued a program with HST/STIS of 40
spiral galaxies of various Hubble types (Rossa et al.~2006). These
spectra have the advantage of higher spatial resolution that can
cleanly separate the NC light from the underlying galaxy
light. However, the $S/N$ and spectral resolution are much lower than
for the VLT spectra, which makes it impossible to infer the NC
kinematics.

We fitted the VLT and the STIS spectra using a non-negative weighted
linear sum of Bruzual \& Charlot (2003) single-age templates with
different ages. This yields for each NC the extinction, metallicity,
star formation history, luminosity-weighted mean log(age), and
$M/L$. Figure~1c shows the STIS spectrum of NGC~300 to illustrate this
approach. The $M/L$ inferred from the spectral population fitting can
be combined with the luminosity inferred from HST imaging to yield the
NC mass. For the nine NCs with VLT spectra, we find that the $M/L$ and
$M$ values inferred from dynamical modeling and spectral population
fitting are in good agreement.

The NCs have mixed-age populations, with single-age fits generally
ruled out by $\chi^2$ statistics. This is supported by an analysis of
line-strength indices measured from the VLT spectra. Approximately
half of the NCs in the STIS sample contain a population younger than
1\,Gyr. The luminosity-weighted ages range from 10\,Myrs to
10\,Gyrs. The NCs in late-type spirals tend to have younger
luminosity-weighted mean ages than those in early-type spirals
(Figure~2b). But even when a young population dominates the light, we
find that there is generally an underlying older population that
contains most of the mass.\looseness=-2

Carollo et al.~(2002) found that NC {\it luminosities} in early-type
spirals are on average brighter than those in late-type spirals. Our
results show that this is true for NC {\it masses} as well: they
correlate loosely with total galaxy luminosity. Figure~2c shows that
there is an even stronger correlation with the luminosity of the host
galaxy {\it bulge}. This correlation has the same slope as the
well-known correlation between supermassive black hole (BH) mass and
bulge luminosity (dashed line in Figure~2c).

\begin{figure}[t]
\includegraphics[height=0.26\hsize]{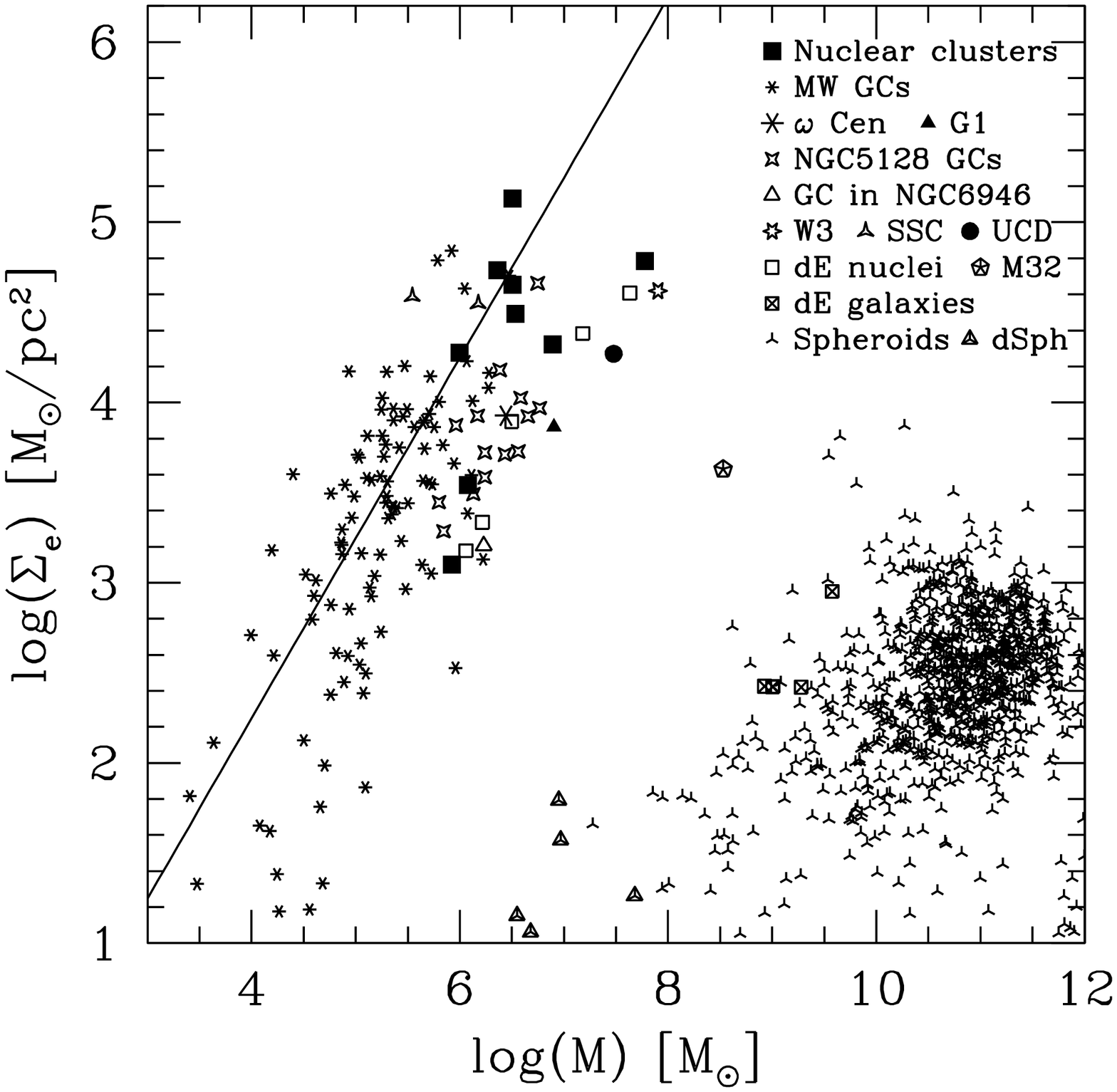}\hfill
\includegraphics[height=0.26\hsize]{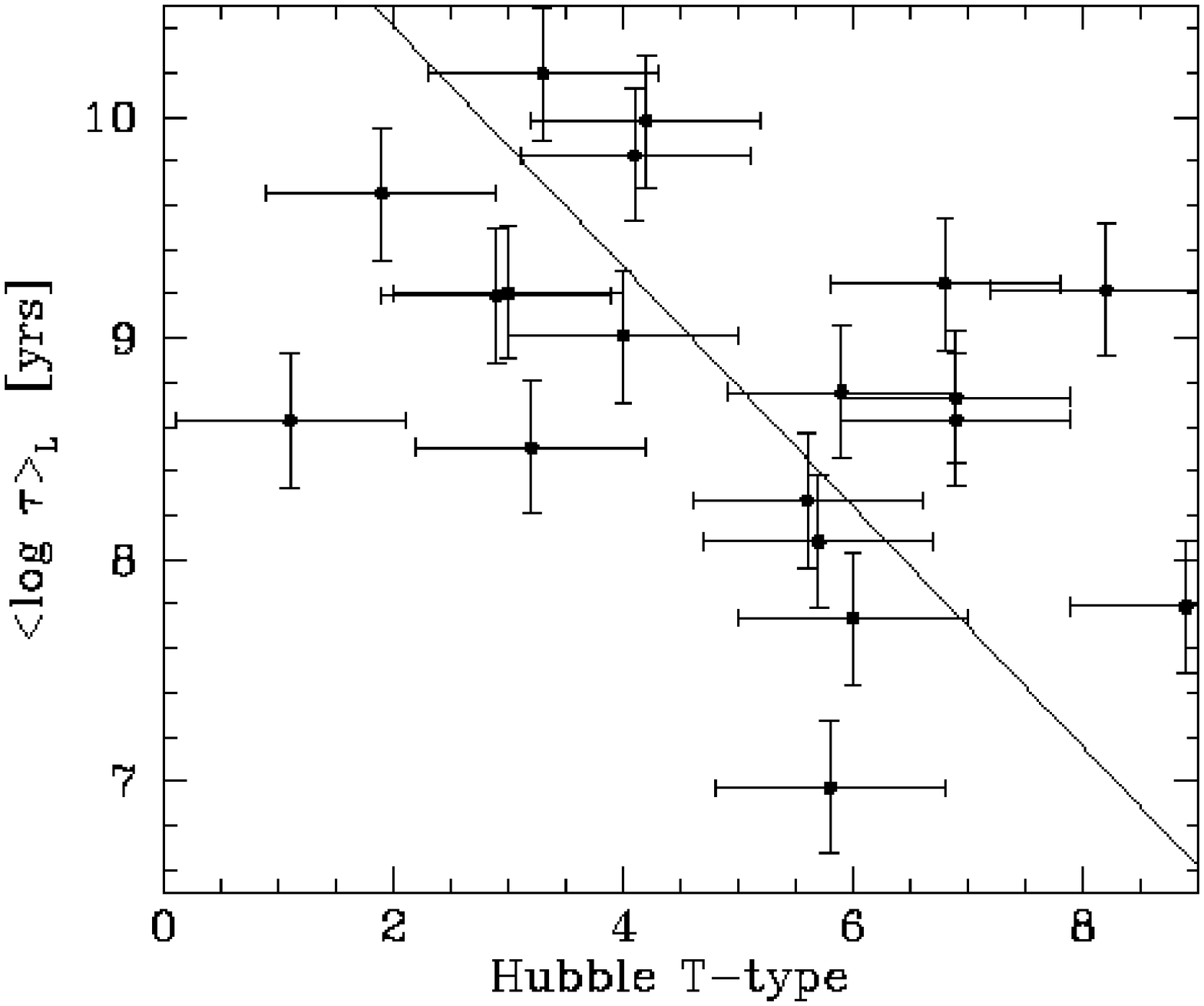}\hfill
\includegraphics[height=0.26\hsize]{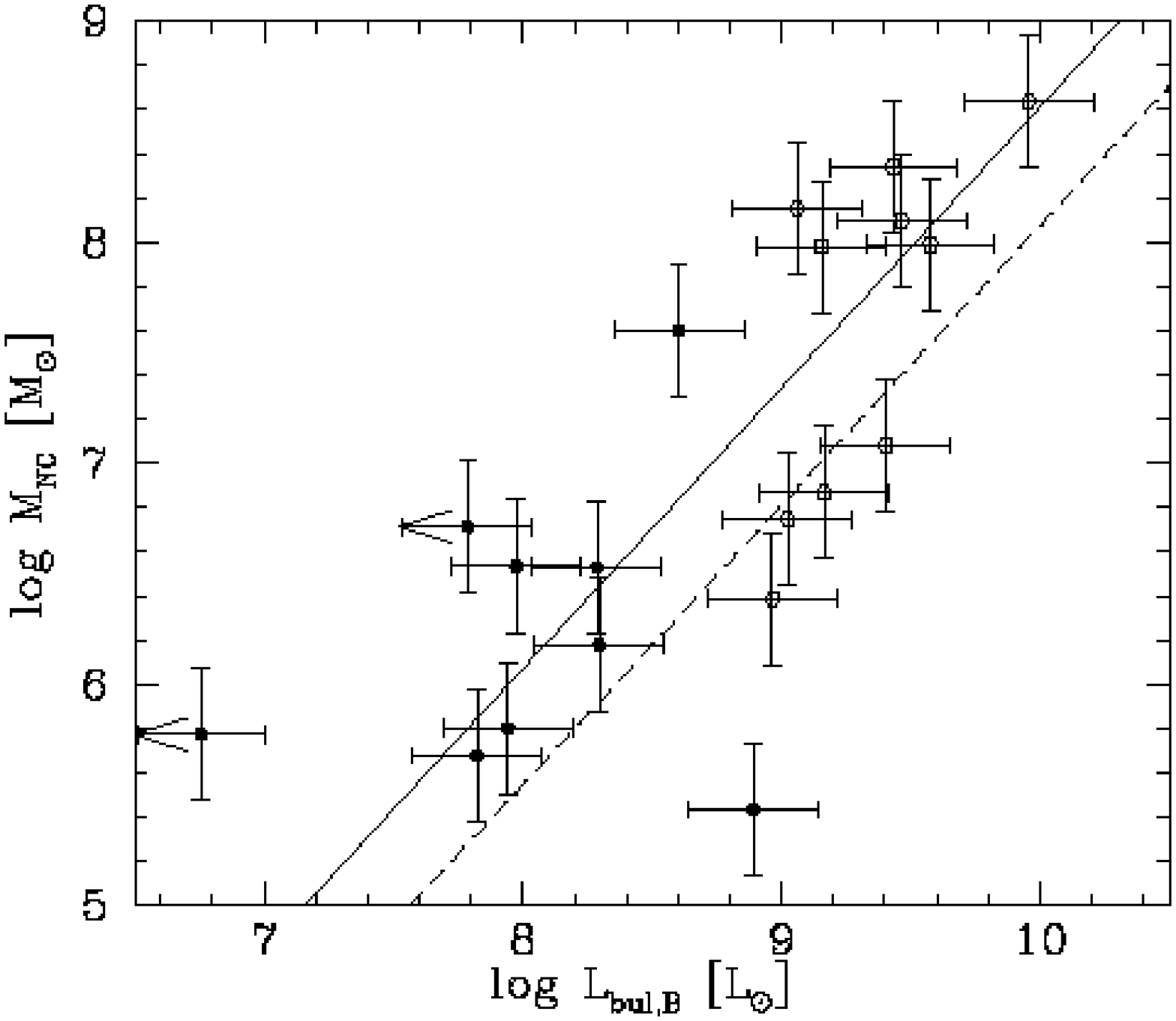}
\caption{(\textit{a}) Mean projected mass density inside $r_{\rm eff}$
versus total mass, for different types of stellar systems as labeled
in the plot (from Walcher et al.~2005). Elliptical galaxies and bulges
are found on the right side of the plot, which presents an
approximately face-on view of their fundamental plane. Globular
clusters are found on the left, roughly along a locus of constant
$r_{\rm eff} \approx 3$ pc (solid line). The nuclear star clusters
(NCs) in spiral galaxies (solid squares) lie at the high-mass end of
the globular cluster sequence, as do the nuclei of dE
galaxies. (\textit{b}) Luminosity-weighted mean $\log ({\rm age/Gyr})$
versus Hubble type.  NCs in late-type spiral galaxies tend to have
younger populations than those in early-type spiral galaxies.
(\textit{c}) NC mass versus $B$-band luminosity of the host galaxy
bulge. Open symbols denote early-type spirals and solid symbols denote
late-type spirals. Galaxies with more luminous bulges tend to have
more massive NCs. Solid lines are least squares fits to the data
points. Panels (b,c) are from Rossa et al.~(2006), and are based on
stellar population fits to HST/STIS spectra such as those shown in
Figure~1c. The dashed line in the right panel indicates the relation
between the supermassive black hole mass and bulge luminosity for
galaxies studied by Marconi \& Hunt (2003).\looseness=-2}
\end{figure}

\section{Discussion}
\noindent The results of the spectroscopic analysis provide insight
into several important questions.

{\bf What do we know about NC formation?} NCs have similar structural
properties as globular clusters. This suggests a commonality in the
physical processes that shaped them, despite their different
environments. By contrast, bulges follow very different scaling
relationships, indicating that their formation was probably governed
by different physical processes. NCs generally have underlying old
stellar populations, suggesting that they may have formed early in the
life of the galaxy. Moreover, NC formation must be efficient, because
NCs/nuclei are a common feature of galaxies. However, there must be
circumstances that prevent NC formation in some fraction of galaxies.

{\bf How do NCs and the galaxies they reside in evolve?} NCs have had
repeated star formation episodes. So they grow in mass as time passes,
due to supply of new gas to the galaxy center through secular
evolution processes (e.g., bar torques). NCs in later-type galaxies on
average have younger populations, probably due to their larger gas
supply. By contrast, nuclei in dEs have predominantly old populations
(Geha, Guhathakurta \& van der Marel 2003), probably reflecting their
general lack of gas. In fact, dEs may have formed through
transformation of late-type spirals (Moore, Lake \& Katz 1998), so dE
nuclei may be the more evolved counterparts of the NCs in spirals. NCs
can also cause secular evolution, given that their masses can be
sufficient to drive the destruction of bars and the formation of
pseudo-bulges (Carollo 1999; Kormendy \& Kennicut 2004).

{\bf What is the relation between NCs and supermassive BHs?} NC masses
scale with host galaxy properties similarly as BH
masses. Contemporaneously with our work on spirals (Rossa et
al.~2006), this was independently shown also for the nuclei in dE
galaxies (Wehner \& Harris 2006) and the nuclei in
intermediate-luminosity elliptical galaxies (C\^ot\'e et al.~2006). In all
cases the NC masses are a few times larger than BH masses that one
would predict based on established relations with $L_{\rm bulge}$
(see, e.g., Figure 2c) or $\sigma$. Ferrarese et al.~(2006a) showed
that better agreement is obtained when host galaxy mass is used as the
independent variable. Also, BHs are typically measured in galaxies
more massive than a few times $10^{10} M_{\odot}$, whereas NCs are
typically detected in less massive galaxies. This latter finding could
indicate that they are different manifestations of a single ``Compact
Massive Object'' class. But it could also be a selection effect; e.g.,
the dwarf Seyfert NGC 4395 has {\it both} a BH and a NC (Filippenko \&
Ho 2003). The implications of all these findings are discussed in
detail in each of the cited papers, including the interesting
possibility that BHs and NCs are somehow causally or evolutionarily
connected. However, this does not necessarily follow from the
data. McLaughlin et al.~(2006) discussed a feedback mechanism that can
account for all observed relations without invoking any direct
connection between BHs and NCs. As of yet, robust insights into the
true BH-NC connection are still mostly missing.\looseness=-2

\begin{acknowledgments}

Support for HST proposals \#9070 and \#9783 was provided by NASA
through a grant from STScI, which is operated by AURA, Inc., under
NASA contract NAS 5-26555.

\end{acknowledgments}

\end{document}